\documentclass[11pt]{JHEP3}

\JHEPspecialurl{http://jhep.sissa.it/JOURNAL/JHEP3.tar.gz}

\usepackage{epsfig,multicol}
\usepackage{amsmath}

\newcommand\fverb{\setbox\fverbbox=\hbox\bgroup\verb}
\newcommand\fverbdo{\egroup\medskip\noindent%
			\fbox{\unhbox\fverbbox}\ }
\newcommand\fverbit{\egroup\item[\fbox{\unhbox\fverbbox}]}
\newbox\fverbbox

\newcommand{\nn}{\nonumber}
\def\dfrac#1#2{\displaystyle\frac{#1}{#2}}

\newcommand{\pslash}{p\kern-1ex /}
\newcommand{\qslash}{q\kern-1ex /}
\newcommand{\lslash}{l\kern-1ex /}
\newcommand{\sslash}{s\kern-1ex /}
\newcommand{\kaslash}{k_a\kern-2ex /}
\newcommand{\kbslash}{k_b\kern-2ex /}
\newcommand{\Dslash}{\mathcal{D}\kern-1.5ex /}

\newcommand{\beqa}{\begin{eqnarray}}
\newcommand{\eeqa}{\end{eqnarray}}
\newcommand{\rmO}{\mathrm{O}}

\newcommand{\ba}{\begin{eqnarray}}
\newcommand{\ea}{\end{eqnarray}}
\newcommand{\be}{\begin{equation}}   
\newcommand{\Nf}{N_\mathrm{f}}

\title{Short distance repulsion in 3 nucleon forces from  perturbative QCD}
\author{Sinya Aoki\\
       Graduate School of Pure and Applied Sciences, University of Tsukuba, Tsukuba, Ibaraki 305-8571,    Japan\\
        E-mail: \email{saoki@het.ph.tsukuba.ac.jp}}

\author{Janos Balog\\
        Research Institute for Particle and Nuclear Physics, 1525 Budapest 114, Pf. 49, Hungary\\
        E-mail: \email{balog@rmki.kfki.hu}}
 \author{Peter Weisz\\
        Max-Planck-Institut f\"ur Physik, F\"ohringer Ring 6, D-80805 M\"unchen, Germany\\
        E-mail: \email{pew@mpp.mpg.de}}       

\received{\today} 		
\accepted{\today}	
\preprint{MPP-2011-120, UTHEP-635}

\abstract{We investigate the short distance behavior of  3 nucleon forces (3NF) defined through Nambu--Bethe--Salpeter wave functions,  using the operator product expansion(OPE)  and calculating anomalous dimensions of 9--quark  operators in perturbative QCD.
As is the case of NN forces previously considered, we show that 3NF have repulsions at short distance at 1--loop, which becomes exact in the short distance limit thanks to the asymptotic freedom of QCD. 
Moreover these behaviors are universal in the sense that they do not depend on the energy of the NBS wave function for 3 nucleons.}


\begin{document}

\section{Introduction} 
Realistic nuclear potentials between two nucleons (2N), determined precisely  
from 2N scattering data together with the deuteron binding energy,
 have often been used to study nuclear many-body problems.  
These two-nucleon forces (2NF), however,
generally underestimate the experimental binding energies of 
light nuclei~\cite{Kamada:2001tv, Pieper:2007ax} and
 this fact indicates the necessity of taking into account three-nucleon forces (3NF). 
 In addition, a clear indication of 3NF is observed in high precision deuteron-proton  elastic scattering data at intermediate energies~\cite{Sekiguchi:2011ku}.

The 3NF may also play an important role for various  phenomena
in nuclear physics and astrophysics, which include
(i) the backward scattering cross sections in 
 nucleus-nucleus elastic scattering~\cite{Furumoto:2009zz},
(ii) the anomaly in the oxygen isotopes near the neutron drip-line~\cite{Otsuka:2009cs}, and 
(iii) the nuclear equation of state at high density  relevant to the physics of neutron stars~\cite{Akmal:1998cf}.
 Universal short distance repulsion for three baryons (nucleons and hyperons) 
 is also suggested to explain the observed maximum mass of 
 neutron stars~\cite{Nishizaki:2002ih}.

Despite of its phenomenological importance,
a microscopic understanding  of 3NF is  still limited, due to difficulties in studying 3NF experimentally. 
Pioneered by Fujita and Miyazawa~\cite{Fujita:1957zz},
the long range part of 3NF has been modeled
 by two-pion exchange~\cite{Coon:2001pv},
which is known to be attractive at long distance.
In addition a repulsive component of 3NF at short distance is 
introduced in a purely phenomenological way~\cite{Pieper:2001ap}.

To go beyond phenomenology, it is most desirable to determine 3NF
directly from the fundamental degrees of freedom , the quarks and the gluons,
on the basis of QCD.
Recently the first investigation of this kind has been attempted using lattice QCD simulations,
where 3NF have been extracted   from the Nambu-Bethe-Salpeter (NBS) wave function for a specific alignment of 3 nucleons \cite{Doi:2010yh,Doi:baryons2010,Doi:2011gq}.  The method used there had been previously employed to extract nucleon-nucleon potentials ({\it i.e.} 2NF)\cite{Ishii:2006ec,Aoki:2008hh,Aoki:2009ji,Ishii:2009zr} as follows.
The NBS wave function for 2 nucleons is defined by
\beqa
\varphi_E(\vec r) &=& \langle 0 \vert N(\vec x+\vec r, t)N(\vec x , t) 
\vert 2{\rm N}, W\rangle\,,
\eeqa
where $\vert 2{\rm N}, W\rangle$ is a QCD eigenstate with energy $W=2\sqrt{m_N^2+k^2}$ with $m_N$ being the nucleon mass, $E=k^2/m_N$ represents the non-relativistic kinetic energy,
and $N$ is a nucleon interpolating operator made of 3 quarks such as 
$N(x) =\epsilon^{abc}q^a(x)q^b(x)q^c(x)$.
The non-local but energy independent potential (more precisely  the half off-shell $T$-matrix) is extracted from this NBS wave function as
\beqa
(E-H_0)\varphi_E(\vec r) &=& \int U(\vec r, \vec {r^\prime}) \varphi_E(\vec {r^\prime}) d^3 r^\prime
\eeqa
where $H_0=-\nabla^2/m_N$.
The non-local potential can be expanded in terms of the velocity (derivative) with local function as
\beqa
U(\vec r, \vec {r^\prime}) = V(\vec r,\vec \nabla)\delta^3(\vec r-\vec {r^\prime}),
\eeqa
which becomes
\beqa
V(\vec r,\vec \nabla) &=& V_0(r) +V_{\sigma}(r)\vec\sigma_1\cdot\vec\sigma_2 + V_T(r) S_{12} + V_{\rm LS}(r) \vec L\cdot \vec S + O(\nabla^2)
\eeqa
 at the lowest few orders,
where $r=\vert \vec r\vert$, $\sigma_i$ represents the Pauli-matrices acting on the spin index of the $i$-th nucleon, $\vec S=(\vec\sigma_1+\vec\sigma_2)/2$ is the total spin, $\vec L =\vec r\times \vec p$ is the angular momentum, and 
\beqa
S_{12} &=& 3\frac{(\vec r\cdot\vec\sigma_1)(\vec r\cdot\vec\sigma_2)}{r^2} -\sigma_1\cdot\sigma_2
\eeqa
is the tensor operator. This method has been shown to work well. The central potentials at the leading order in the expansion  have qualitatively reproduced common features of phenomenological 2N potentials:
the force is attractive at medium to long distance while it has a characteristic repulsive core at short distance. See also refs.~\cite{aoki_review,aoki_lecture} for  a summary of results and recent developments.

The present authors have investigated short distance behaviors of the 2NF defined in the framework  mentioned above, using the operator product expansion(OPE) and
perturbation theory thanks to asymptotic freedom of QCD
\cite{Aoki:2009pi,Aoki:2010kx,Aoki:2010uz} . (See also a similar attempt for the solvable models in 2 dimensions \cite{Aoki:2008yw}.)
The behavior of the NBS wave function $\varphi_E(\vec r)$ at short distance ($r\rightarrow 0$)
is encoded in the operator product expansion (OPE) of 
the two nucleon operators: 
\begin{equation}
N(\vec x/2,0)\,N(-\vec x/2,0)\approx\sum_k
D_k(\vec x)\,{\cal O}_k(\vec 0,0),
\end{equation}
where $\{ {\cal O}_k \}$ is a set of local color singlet 6-quark operators 
with two-nucleon quantum numbers. Asymptotically the $\vec x$-dependence 
and energy dependence of the wave function is factorized into
\begin{equation}
\varphi_E(\vec x)\approx\sum_k D_k(\vec x)
\langle0\vert{\cal O}_k(\vec 0,0)\vert 2{\rm N},W\rangle\,.
\label{eq:wf}
\end{equation}
Standard renormalization group (RG) analysis  gives \cite{Aoki:2010kx} the leading 
short distance behavior of the OPE coefficient function as
\begin{equation}
D_k(\vec x)\approx \left(\ln\frac{r_0}{r}\right)^{\nu_k}\,d_k\,,
\end{equation}
where $\nu_k$ is related to the 1-loop coefficient of the anomalous dimension
of the operator ${\cal O}_k$, $d_k$ is the tree-level 
contribution 
of $D_k(\vec 0)$ and finally $r_0$ is some typical non-perturbative
QCD scale.
Assuming its matrix element does not vanish, the operator with largest
RG power $\nu_k$ dominates the wave function (\ref{eq:wf}) at short distances.
We denote the largest power by $\nu_1$ and the second largest one by $\nu_2$.

If $\nu_1$ is non-zero, this leads to the leading
asymptotics of the s-wave potential of the form
\begin{equation}
V(r)\approx-\frac{\nu_1}{r^2\left(\ln\frac{r_0}{r}\right)}\,,
\end{equation}
which is attractive for $\nu_1>0$ and repulsive for $\nu_1<0$.

If $\nu_1=0$, the situation is more complicated.
The relative sign of the ratio $R$ between the leading and the subleading
contributions becomes important and we find:
\begin{equation}
V(r)\approx-R\frac{\nu_2}{r^2\left(\ln\frac{r_0}{r}\right)^{1-\nu_2}}\,.
\label{nu2}
\end{equation}
If $R$ is positive, the potential is
repulsive, while it is attractive for negative $R$. A system of two nucleons
corresponds to this degenerate case. Unfortunately in this case $R$ depends
on the energy $E$. In \cite{Aoki:2010kx}
it is argued that in the relevant energy range the relative coefficient $R$
is positive, so that the
short distance limit of the nucleon potential is repulsive.

In this paper, we extend the above OPE analysis to the 3NF.
The corresponding equal time NBS wave function for 3 nucleons is given by
\beqa
\psi_{3N}(\vec r, \vec \rho) &=& \langle 0 \vert N(\vec x_1,0) N(\vec x_2,0) N(\vec x_3,0)\vert E_{3N}\rangle
\eeqa
where $E_{3N}$ and $\vert E_{3N}\rangle$ denote the energy and the 3N state. We introduce Jacobi coordinates  $\vec R =(\vec x_1+\vec x_2+\vec x_3)/3$, $\vec r=(\vec x_1-\vec x_2)/2$, $\vec \rho = (\vec x_3 -(\vec x_1+\vec x_2)/2)/\sqrt{3} $.
From this wave function, the three nucleon potential is defined by
\beqa
\left[ -\frac{1}{2\mu_r}\nabla_r^2 -\frac{1}{2\mu_\rho}\nabla^2_\rho +\sum_{i< j} V_{2N}(\vec{r}_{ij})
+ V_{3NF}(\vec r,\vec \rho)\right] \psi_{3N}(\vec r,\vec \rho) = E_{3N} \psi_{3N}(\vec r,\vec\rho)
\eeqa
where $V_{2N}(\vec{r}_{ij})$ with $\vec{r}_{ij} =\vec x_i -\vec x_j$ denotes 2NF between $(i,j)$-pair, $V_{3NF}(\vec r,\vec \rho)$ the 3NF, $\mu_r =\mu_\rho = m_N/2$ the reduced masses.

In sect.~\ref{sec:RG}, we start with renormalization group 
considerations and OPE, which are relevant for 3NF. The anomalous dimensions of 9--quark operators
are computed in sect.~\ref{sec:result}. Finally  we discuss the short distance behavior of 3NF in sect.~\ref{sec:discussion}. 
For the convenience of the reader we give a 
brief summary of our results here.
The OPE analysis shows that the 3N central potential at short distance behaves 
as
\beqa
V_{3NF}(\vec r,\vec\rho) \simeq \frac{ 1}{m_N} \frac{-4\beta_A^{\rm tree}}{s^2(-\log(\Lambda s))} ,
\eeqa
as $s=\sqrt{(\vec r)^2 +(\vec \rho)^2}\rightarrow 0$,
where $\beta_A^{\rm tree}$ is given by
\beqa
\beta_A^{\rm tree} = -14/(33-2N_f) \, , 
\label{eq:betaA}
\eeqa
where $N_f$ is the number of dynamical quarks.
Unlike the 2NF where  the situation was not completely determined by PT alone,
it is shown that the 3N potential always has a repulsive core.
Furthermore  it is universal in the sense that it does not depend on details of the 3N state used to define the NBS  wave function such as the energy of the state.

\section{Renormalization group analysis and operator product expansion for 3NF}
\label{sec:RG}

\subsection{Renormalization group equation for composite operators}

In QCD, using dimensional regularization in $D=4-2\epsilon$ dimensions,
bare local composite operators $O^{(0)}_A(x)$ are renormalized
according to
\begin{equation}
O^{({\rm ren})}_A(x)=Z_{AB}(g,\epsilon)\,O^{(0)}_B(x)\,.
\end{equation}
Summation of repeated indices is assumed throughout this paper unless
indicated otherwise.
The meaning of the above formula is that we obtain finite results
if we insert the right hand side into any correlation function of the
fundamental gluon and quark fields, provided
we also renormalize the bare QCD coupling $g_0$ and the quark and gluon fields.
For example, in the case of an $n$--quark correlation function with operator
insertion, which we denote by ${\cal G}^{(0)}_{n;A}(g_0,\epsilon)$ 
(suppressing the dependence on the quark momenta and other quantum numbers)
we have
\begin{equation}
{\cal G}^{({\rm ren})}_{n;A}(g,\mu)=Z_{AB}(g,\epsilon)\,
Z_F^{-n/2}(g,\epsilon)\, 
{\cal G}^{(0)}_{n;B}(g_0,\epsilon)\,. 
\end{equation}
We recall from renormalization theory
that for the analogous $n$--quark 
correlation function (without any operator insertion) we have
\begin{equation}
{\cal G}^{({\rm ren})}_n(g,\mu)=Z_F^{-n/2}(g,\epsilon)\, 
{\cal G}^{(0)}_n(g_0,\epsilon)\, .
\end{equation}
The coupling renormalization is given by
\begin{equation}
g_0^2=\mu^{2\epsilon}\,Z_1(g,\epsilon)\,g^2\,.
\end{equation}
The renormalization constant $Z_1$ in the minimal subtraction (MS)
scheme we are using has pure pole terms only:
\begin{equation}
Z_1(g,\epsilon)=1-\frac{\beta_0g^2}{\epsilon}-\frac{\beta_1g^4}{2\epsilon}
+\frac{\beta_0^2g^4}{\epsilon^2}+\rmO(g^6)\,,
\end{equation}
where
\begin{equation}
\beta_0=\frac{1}{16\pi^2}\left\{\frac{11}{3}N_c-\frac{2}{3}N_f\right\}\,,
\qquad\quad
\beta_1=\frac{1}{256\pi^4}\left\{\frac{34}{3}N_c^2-\left(
\frac{13}{3}N_c-\frac{1}{N_c}\right)N_f\right\}\,.
\end{equation}
Similarly for the fermion field renormalization constant, we have
\begin{equation}
Z_F(g,\epsilon)=1-\frac{\gamma_{F0}g^2}{2\epsilon}+\rmO(g^4)\,,
\end{equation}
where $\gamma_{F0}=\dfrac{\lambda C_F}{8\pi^2}$ with $C_F=\dfrac{N_c^2-1}{2N_c}$ and $\lambda$ is the covariant gauge parameter.
The gluon field renormalization constant is also similar, but we do not need 
it here. Finally the matrix of operator renormalization constants
is of the form
\begin{equation}
Z_{AB}(g,\epsilon)=\delta_{AB}-\frac{\gamma^{(1)}_{AB}g^2}{2\epsilon}
+\rmO(g^4)\,.
\end{equation}

The renormalization group (RG) expresses the simple fact that bare quantities
are independent of the renormalization scale $\mu$. Introducing the RG 
differential operator
\begin{equation}
{\cal D}=\mu\frac{\partial}{\partial\mu}+\beta(g)\,\frac{\partial}{\partial g}
\end{equation}
the RG equation for $n$--quark correlation functions can be written as
\begin{equation}
\left\{{\cal D}+\frac{n}{2}\gamma_F(g)\right\}\,
{\cal G}^{({\rm ren})}_n(g,\mu)=0\, .
\end{equation}
 Here the RG beta function is
\begin{equation}
\beta(g)=\epsilon g +\beta_D(g,\epsilon)
=\epsilon g-\frac{\epsilon g}{1+\frac{g}{2}\,
\frac{\partial\ln Z_1}{\partial g}}=-\beta_0g^3-\beta_1g^5+\rmO(g^7)\,,
\end{equation}
where $\beta_D(g,\epsilon)$ is the beta function in $D$ dimensions
and the RG gamma function (for quark fields) is
\begin{equation}
\gamma_F(g)=\beta_D(g,\epsilon)\,\frac{\partial\ln Z_F}
{\partial g}=\gamma_{F0}\,g^2+\rmO(g^4)\,.
\end{equation}
It is useful to introduce the RG invariant lambda-parameter $\Lambda$
by taking the ansatz 
\begin{equation}
\Lambda=\mu\,{\rm e}^{f(g)}
\end{equation}
and requiring ${\cal D}\Lambda=0$. The solution is the lambda-parameter
in the MS scheme ($\Lambda_{\rm MS}$) if the arbitrary integration constant
is fixed by requiring that for small coupling
\begin{equation}
f(g)=-\frac{1}{2\beta_0g^2}-\frac{\beta_1}{2\beta_0^2}\,\ln(\beta_0g^2)
+\rmO(g^2)\,.
\end{equation}
Finally the RG equations for $n$--quark correlation functions with
operator insertion are of the form
\begin{equation}
\left\{{\cal D}+\frac{n}{2}\gamma_F(g)\right\}\,
{\cal G}^{({\rm ren})}_{n;A}(g,\mu)-\gamma_{AB}(g)
{\cal G}^{({\rm ren})}_{n;B}(g,\mu)=0\,,
\end{equation}
where
\begin{equation}
\gamma_{AB}(g)=-Z_{AC}\beta_D(g,\epsilon)\frac
{\partial Z^{-1}_{CB}}{\partial g}=\gamma^{(1)}_{AB}g^2+\rmO(g^4)\,.
\end{equation}

\subsection{OPE and RG equations}

Let us recall the operator product expansion for $\rho,r\to0$:
\begin{equation}
O_1(\vec r-\vec\rho/\sqrt{3})O_2(-\vec r-\vec \rho/\sqrt{3})O_3(2\vec\rho/\sqrt{3}) \simeq D_B(\vec r,\vec \rho)\,O_B(0)\,.
\label{ope33}
\end{equation}
We will need it in the special case where the operators $O_1,O_2,O_3$ on the
left hand side are nucleon operators and the set of operators $O_B$ on
the right hand side are local 9-quark operators of engineering dimension 27/2
and higher. All operators in (\ref{ope33}) are renormalized ones, but from 
now on we suppress the labels $^{({\rm ren})}$.
As we know, the nucleon operators are renormalized diagonally as
\begin{equation}
O_i=\zeta_i(g,\epsilon)\,O^{(0)}_i\,,
\end{equation}
and we can define the corresponding RG gamma functions by
\begin{equation}
\gamma_{i}(g)=\beta_D(g,\epsilon)\,
\frac{\partial\ln\zeta_{i}}{\partial g}=\gamma_{i}^{(1)}g^2+\rmO(g^4)\,.
\end{equation}
For the nucleon operator, 
\beqa
\gamma^{(1)}_N &=& 24 d, \quad  d=\frac{1}{32 N_c\pi^2} =\frac{1}{96\pi^2} .
\eeqa

Next we write down the bare version of (\ref{ope33}) (in terms of bare 
operators and bare coefficient functions):
\begin{equation}
O_1^{(0)}(\vec r-\vec\rho/\sqrt{3})O_2^{(0)}(-\vec r-\vec\rho/\sqrt{3})O_3^{(0)}(2\vec\rho/\sqrt{3}) \simeq D_B^{(0)}(\vec r,\vec\rho)\,O_B^{(0)}(0)\,.
\label{ope33bare}
\end{equation}
Comparing (\ref{ope33}) to (\ref{ope33bare}), we can read off the 
renormalization of the coefficient functions:
\begin{equation}
D_B(\vec r,\vec \rho)=\zeta_1(g,\epsilon)\zeta_2(g,\epsilon)\zeta_3(g,\epsilon)D^{(0)}_A(\vec r,\vec\rho)\,
Z^{-1}_{AB}(g,\epsilon)
\end{equation}
and using the $\mu$-independence of the bare coefficient functions
we can derive the RG equations satisfied by the renormalized ones:
\begin{equation}
{\cal D}D_B(g,\mu,\vec r,\vec\rho)+D_A(g,\mu,\vec r,\vec \rho)\,\tilde\gamma_{AB}(g)=0\,,
\label{RG33}
\end{equation}
where the effective gamma function matrix is defined as
\begin{equation}
\tilde\gamma_{AB}(g)=\gamma_{AB}(g)-\left[\gamma_1(g)+\gamma_2(g)+\gamma_3(g)\right]
\,\delta_{AB}\,.
\end{equation}

\subsection{Perturbative solution of the RG equation and factorization
of OPE}

We want to solve the vector partial differential equation (\ref{RG33}) and
for this purpose it is useful to introduce $\hat U_{AB}(g)$,
the solution of the matrix ordinary differential equation
\begin{equation}
\beta(g)\,\frac{{\rm d}}{{\rm d}g}\,\hat U_{AB}(g)=\tilde\gamma_{AC}(g)
\,\hat U_{CB}(g)
\label{hatU}
\end{equation}
and its matrix inverse $U_{AB}(g)$. We will assume that the coefficient
has the perturbative expansion
\beqa
D_A(g,\mu,\vec r,\vec\rho)&=&\sum_{\alpha_1+\alpha_2=\tilde d_A} r^{\alpha_1}\rho^{\alpha_2} D_A^{\alpha_1\alpha_2}(g;\mu s, \omega, \Omega_r, \Omega_\rho)\nn \\
&=& \sum_{\alpha_1+\alpha_2=\tilde d_A} r^{\alpha_1}\rho^{\alpha_2}\left[D_{A;0}^{\alpha_1\alpha_2}+g^2D_{A;1}^{\alpha_1\alpha_2}(\mu s, \omega, \Omega_r,\Omega_\rho)+\rmO(g^4)\right]\,,
\label{Dpert}
\eeqa
where $s^2=r^2+\rho^2$ and $\tan\omega=\rho/r$ with $r=\vert \vec r\vert$, $\rho=\vert\vec \rho\vert$,  and $\Omega_r, \Omega_\rho$ are solid angles of the vectors $\vec r$ and $\vec\rho$, respectively. Here $\tilde d_A = d_A -(d_1+d_2+d_3)$ is the dimension of the coefficient function.
Note that in the massless theory operators of different dimension
do not mix.
In the full theory quark mass terms are also present, but they
correspond to higher powers in $r$  and $\rho$, and therefore can be neglected.

We will also assume that the basis of operators has been chosen such that the
1-loop mixing matrix is diagonal:
\begin{equation}
\tilde\gamma_{AB}(g)=2\beta_0\,\beta_A\,g^2\,\delta_{AB}+\rmO(g^4)\,.
\end{equation}
In such a basis the solution of (\ref{hatU}) in perturbation theory
takes the form
\begin{equation}
\hat U_{AB}(g)=\left\{\delta_{AB}+R_{AB}(g)\right\}\,g^{-2\beta_B}\,,
\label{hatUpert}
\end{equation}
where $R_{AB}(g)=\rmO(g^2)$, with possible multiplicative $\log g^2$ factors,
depending on the details of the spectrum of 1-loop eigenvalues $\beta_A$.

Having solved (\ref{hatU}) we can write down the most general solution
of (\ref{RG33}):
\begin{equation}
D_B^{\alpha_1\alpha_2}(g;\mu s, \omega,\Omega_r,\Omega_\rho)=F_A^{\alpha_1\alpha_2}(\Lambda s,\omega, \Omega_r,\Omega_\rho)\,U_{AB}(g)\,.
\end{equation}
Here the vector $F_A^{\alpha_1\alpha_2}$ is RG-invariant. Introducing the running
coupling $\bar g$ as the solution of the equation
\begin{equation}
f(\bar g)=f(g)+\ln(\mu s)=\ln(\Lambda s)
\end{equation}
$F_B^{\alpha_1\alpha_2}$ can be rewritten as
\begin{equation}
F_B^{\alpha_1\alpha_2}(\Lambda s,\omega, \Omega_r,\Omega_\rho)=D_A^{\alpha_1\alpha_2}(\bar g;1,\omega,\Omega_r,\Omega_\rho)\,\hat U_{AB}(\bar g)\,.
\end{equation}
Since, because of asymptotic freedom (AF), for $s\to0$ also $\bar g\to0$ as
\begin{equation}
\bar g^2\approx-\frac{1}{2\beta_0\ln(\Lambda s)}\,,
\end{equation}
$F_B^{\alpha_1\alpha_2}$ can be calculated perturbatively using (\ref{Dpert}) and 
(\ref{hatUpert}).

Putting everything together, we find that the right hand side of the
operator product expansion (\ref{ope33}) can be rewritten:
\beqa
O_1(\vec r-\vec\rho/\sqrt{3})O_2(-\vec r-\vec \rho/\sqrt{3})O_3(2\vec\rho/\sqrt{3}) &\simeq& 
\sum_{\alpha_1+\alpha_2=\tilde d_B}r^{\alpha_1}\, \rho^{\alpha_2}  \nn \\
&\times& F_B^{\alpha_1\alpha_2}(\Lambda s,\omega, \Omega_r,\Omega_\rho) \,\tilde O_B(0)\,.
\label{ope34}
\eeqa
where
\begin{equation}
\tilde O_B=U_{BC}(g)\,O_C\,.
\end{equation}
There is a factorization of the operator product into perturbative and 
non-perturbative quantities: $F_B^{\alpha_1\alpha_2}(\Lambda s,\omega,\Omega_r,\Omega_\rho)$ is perturbative and 
calculable (for $s \to0$) thanks to AF, whereas the matrix elements of
$\tilde O_B$ are non-perturbative (but $s$-independent). Note that $d_C=d_B$.

\subsection{3NF from OPE}
Using results in the previous subsection, the NBS wave function for 3N can be written at short distance as
\begin{eqnarray}
\psi_{3N}(\vec r,\vec \rho) &\simeq& \sum_{A,B} \sum_{\alpha_1+\alpha_2=\tilde d_A}r^{\alpha_1} \rho^{\alpha_2}\, D_A^{\alpha_1\alpha_2}(\bar g, 1, \omega,\Omega_r,\Omega_\rho) \hat U_{AB}(\bar g) \langle 0 \vert \tilde O_B(0)\vert E_{3N}\rangle . 
\end{eqnarray}
Since a $r^{\alpha_1}\rho^{\alpha_2}$ term produces angular momenta $l_1\le \alpha_1$ and $l_2\le\alpha_2$, we can write
\beqa
D_A^{\alpha_1\alpha_2}(\bar g, 1, \omega,\Omega_r,\Omega_\rho) &\simeq&  \sum_{m_1m_2}D_A^{\alpha_1m_1\alpha_2 m_2}(\bar g,1,\omega) Y_{\alpha_1m_1}(\Omega_r)Y_{\alpha_2m_2}(\Omega_\rho) ,
\eeqa
up to less singular terms at short distances.
Then the sum of 2N and 3NF potentials $V_{2N+3NF}$ is extracted as
\beqa
V_{2N+3NF}(\vec r,\vec\rho)&\equiv&
\sum_{i<j} V_{2N}(\vec r_{ij}) + V_{3NF}(\vec r,\vec\rho)
= E_{3N} +\frac{1}{m_N}\frac{(\nabla^2_r+\nabla_\rho^2)\psi_{3N}(\vec r,\vec\rho)}{  \psi_{3N}(\vec r,\vec\rho)},
\eeqa 
where 
\beqa
\nabla^2_r &=& \frac{1}{r^2} \frac{\partial}{\partial r} r^2  \frac{\partial}{\partial r} -\frac{\hat L_r^2}{r^2}\equiv d_r^2 -\frac{\hat L_r^2}{r^2},\quad
\nabla^2_\rho = \frac{1}{\rho^2} \frac{\partial}{\partial \rho} \rho^2  \frac{\partial}{\partial \rho} -\frac{\hat L_\rho^2}{\rho^2}\equiv d_\rho^2 -\frac{\hat L_\rho^2}{\rho^2} .
\eeqa
Since 
\beqa
\nabla^2_r  r^l Y_{lm}(\Omega_r) = \nabla_\rho^2 \rho^l Y_{lm} (\Omega_\rho)&=& 0,
\eeqa
we have
\beqa
(\nabla^2_r+\nabla_\rho^2)\psi_{3N}(\vec r,\vec\rho) &\simeq& \sum_{AB}\sum_{\alpha_1+\alpha_2=\tilde d_A}r^{\alpha_1} Y_{\alpha_1m_1}(\Omega_r) \rho^{\alpha_2} Y_{\alpha_2m_2}(\Omega_\rho)  \langle 0 \vert \bar O_B(0)\vert E_{3N}\rangle \nn \\
&\times& (d_r^2+d_\rho^2)\left[ D_A^{\alpha_1m_1\alpha_2 m_2}(\bar g,1,\omega) \hat U_{AB}(\bar g)\right] .
\eeqa
Since terms with $\tilde d_A=0$ dominate in $\Psi_{3N}$ at short distance, contributions from $\tilde d_A\not=0$ terms to 2N+3NF potentials are  suppressed 
by an $r^{\alpha_1}\rho^{\alpha_2}$ factor, so that they do not contribute at short distance. Therefore we consider terms with $\tilde d_A=0$ ($\alpha_1=\alpha_2=0$)  hereafter and do not write $\alpha_i$ dependence in coefficients.
We then have
\beqa
(\nabla^2_r+\nabla_\rho^2)\psi_{3N}(\vec r,\vec\rho) &\simeq& \sum_{AB}\langle 0 \vert \bar O_B(0)\vert E_{3N}\rangle
\times (d_r^2+d_\rho^2)\left[ D_A(\bar g,1,\omega) \hat U_{AB}(\bar g)\right] .
\eeqa  

In terms of $s$ and $\omega$ we write
\beqa
d_r^2 +d_\rho^2 = \frac{1}{s^5}\frac{\partial}{\partial s} s^5 \frac{\partial}{\partial s} +\frac{1}{s^2}\left[
 \frac{\partial^2}{\partial\omega^2} +4\frac{\cos(2\omega)}{\sin(2\omega)}\frac{\partial}{\partial\omega}
 \right],
\eeqa
so that the $\omega$ dependent part gives a $1/s^2$ contribution unless $\omega=0,\pi/2,\pi,3\pi/2$, where
either $r=0$ or $\rho=0$. We assume $r\not=0$ and $\rho\not=0$ in our calculation.
Since an $\omega$ dependence appears only at 1-loop or higher orders in $D_A$, we can neglect it unless an operator $O_A$ which appears at $\ell_A$ loop has large anomalous dimension such that $\beta_A-\ell_A$ is larger than other $\beta_B$ corresponding  to operators $O_B$ appearing at tree level. As we will see later such operators are absent; it is then enough to consider the tree level contribution in $D_A$, so that $\omega$ dependent terms in  $D_A$ can be neglected. 
The largest eigenvalue among operators appearing at tree level  is thus denoted by $\beta_A$,
which corresponds to $\nu_1$ discussed in the introduction for 2N forces. 

We then obtain
\beqa
(d_r^2+d_\rho^2)\left[ D_A(\bar g,1,\omega) \hat U_{AB}(\bar g)\right]
&\simeq& D_{A:0}\, (d_r^2+d_\rho^2)\,  \bar g^{-2\beta_A} \nn \\
&\simeq&  D_{A:0}\frac{-4\beta_A}{s^2(-\log(\Lambda s))}(-2\beta_0\log(\Lambda s))^{\beta_A} .
\eeqa
The NBS wave function is dominated by the term with largest $\beta_A$. If we assume 
$\beta_A$ is non-zero, we finally obtain
\beqa
V_{2N+3NF}(\vec r,\vec\rho) \simeq \frac{ 1}{m_N} \frac{-4\beta_A}{s^2(-\log(\Lambda s))} .
\eeqa


\section{Anomalous dimensions for three nucleons at 1-loop}
\label{sec:result}

\subsection{OPE for 3N operators at tree level}
The general form of a gauge invariant local 3--quark operator is given by
\beqa
B^F_\Gamma (x) \equiv B^{fgh}_{\alpha\beta\gamma}(x) 
= \varepsilon^{abc} q^{a,f}_\alpha(x) q^{b,g}_\beta(x) q^{c,h}_\gamma (x)\,,
\label{baryonop}
\eeqa
where $\alpha,\beta,\gamma$ are spinor, $f,g,h$ are flavor, 
$a,b,c$ are color indices of the (renormalized) quark field $q$.  
The color index runs from 1 to $N_c=3$, the spinor index from 1 to 4, 
and the flavor index from 1 to $\Nf$. 
Note that $B^{fgh}_{\alpha\beta\gamma}$ is symmetric under any
interchange of pairs of indices 
(e.g. $B^{fgh}_{\alpha\beta\gamma}=B^{gfh}_{\beta\alpha\gamma}$)
because the quark fields anticommute.
For simplicity we sometimes use the 
notation such as $F=fgh$ and $\Gamma=\alpha\beta\gamma$ as indicated in
(\ref{baryonop}).

The usual nucleon operator which is employed in lattice simulations
is constructed from the above operators as
\beqa
B^f_\alpha(x) = \left(P_{+4}\right)_{\alpha\alpha'} 
B_{\alpha'\beta\gamma}^{fgh} (C\gamma_5)_{\beta\gamma}(i\tau_2)^{gh}\,,
\eeqa 
where $P_{+4} = (1+\gamma_4)/2$ is the projection to the large 
spinor component, $C=\gamma_2\gamma_4$ is the charge conjugation matrix, 
and $\tau_2$ is the Pauli matrix in the flavor space (for $\Nf=2$) given by 
$(i\tau_2)^{fg} = \varepsilon^{fg}$. 
Both $C\gamma_5$ and $i\tau_2$ are anti-symmetric under
the interchange of two indices, so that the nucleon operator 
has spin $1/2$ and isospin $1/2$.  Although an explicit form 
of the $\gamma$ matrices is unnecessary in principle, 
we find it convenient to use a (chiral) convention given by
\beqa
 \gamma_k &=& \left( \begin{array}{cc}
 0 & i\sigma_k \\
 -i\sigma_k & 0
 \end{array}
 \right)\,, \
 \gamma_4 = \left( \begin{array}{cc}
 0 & {\bf 1} \\
 {\bf 1} & 0 \\
 \end{array}
 \right)\,, \
 \gamma_5 =\gamma_1\gamma_2\gamma_3\gamma_4 =
 \left( \begin{array}{cc}
 {\bf 1}& 0  \\
 0 & -{\bf 1} \\
 \end{array}
 \right)\,.
 \eeqa

As discussed in the previous section, the OPE at the 
tree level (generically) dominates at short distance.
The OPE of 3N operators given above at tree level becomes 
\beqa
B^f_{\alpha}(x+y-z/\sqrt{3}) B^g_{\beta}(x-y-z/\sqrt{3})B^h_{\gamma}(x+2z/\sqrt{3}) &=& B^f_{\alpha}(x) B^g_{\beta}(x)B^h_{\gamma}(x)+\cdots 
\eeqa
where $+\cdots$ denote higher dimensional operators, which do not contribute at short distance.
The leading operators couple only to  the states with $L=0$  (the $S$-state).

If we construct local 3N operators at $L=0$ from $B^f_\alpha(x)$ for nucleons (which only involve 2 different flavors), there is only one such operator, which has $I=1/2$ and $S=1/2$, due to the Pauli statistics of nucleons.
Explicitly it is given by
\beqa
(B^{(3)}_{\rm tree})^{ffg}_{\alpha\beta\alpha}\equiv B^f_\alpha B^f_\beta B^g_\alpha, \quad B^f_\alpha = B_{\alpha+\hat\alpha, [2,1] +[\hat 2,\hat 1]}^{ffg}
\label{eq:local3N}
\eeqa
where $f\not= g$, $f,g=u,d$ and  $\alpha\not=\beta$, $\alpha,\beta =1,2$, $\hat 1= 3$, $\hat 2= 4$ for the explicit form of the $\gamma$ matrices. Above no summation is taken for $f$ and $\alpha$.

\subsection{General formula for the divergent part at 1-loop}
As shown in ref.\cite{Aoki:2010kx}, the gauge invariant part  of the divergence from diagrams involving exchange of a gluon between any pair of quark fields is given by
\beqa
\left[ q^{a,f}_\alpha(x) q^{b,g}_\beta (x)\right]^{\rm 1-loop, div}
&=& \frac{g^2d}{\epsilon} 
\left[ {\bf T}_0  \cdot 
q^a(x)\otimes q^b (x)\right]_{\alpha,\beta}^{fg} ,
\label{eq:1-loopT}
\eeqa
where
\beqa
({\bf T}_0)^{f f_1,g g_1}_{\alpha\alpha_1,\beta\beta_1} &=& 
\delta^{ff_1}\delta^{gg_1}
\left[ \delta_{\alpha\alpha_1}\delta_{\beta\beta_1} 
    - 2\delta_{\beta\alpha_1}\delta_{\alpha\beta_1}\right]
+N_c\delta^{gf_1}\delta^{fg_1}
\left[ \delta_{\beta\alpha_1}\delta_{\alpha\beta_1}
-2\delta_{\alpha\alpha_1}\delta_{\beta\beta_1} \right] \nn \\
\label{eq:T0}
\eeqa
for either $\alpha_1,\beta_1 \in \{1,2\}$(right-handed) or $\alpha_1,\beta_1 \in \{3,4\}$(left-handed),
and it vanishes for other combinations. 

The operator in eq. (\ref{eq:local3N}) can be written as a linear combination of simple operators $[BBB]^{F_1F_2F_3}_{\Gamma_1\Gamma_2\Gamma_3}$.
According to this 1-loop formula for divergences,
such a simple operator mixes only with operators $[BBB]^{F_AF_BF_C}_{\Gamma_A\Gamma_B\Gamma_C} $ which preserve the set of flavor and Dirac indices in the chiral basis as
\beqa
F_1\cup F_2\cup F_3&=&F_A\cup F_B\cup F_C, \quad
\Gamma_1\cup \Gamma_2\cup \Gamma_3=\Gamma_A\cup \Gamma_B\cup \Gamma_C.
\eeqa
Note however that such operators are not all linearly independent.
In the case of a 2N operator, we have the following constraint
\beqa
3 [BB]^{F_1,F_2}_{\Gamma_1,\Gamma_2} + \sum_{i,j=1}^3 [BB]^{(F_1,F_2)[i,j]}_{(\Gamma_1,\Gamma_2)[i,j]} = 0,
\label{gaugeid2}
\eeqa
which comes from the general identity
\beqa
N_c \varepsilon^{a_1\cdots a_{N_c}}\varepsilon^{b_1\cdots b_{N_c}} &=&\sum_{j,k=1}^{N_c}
\varepsilon^{a_1\cdots a_{j-1}b_k a_{j+1}\cdots a_{N_c}}\varepsilon^{b_1\cdots b_{k-1}a_j b_{k+1}\cdots b_{N_c}} .
\eeqa
Here $[i,j]$ means a simultaneous exchange between the $i$-th indices of $F_1, \Gamma_1$ and the $j$-th indices of  $F_2, \Gamma_2$.
This identity can be generalized to
\beqa
N_c\,  \varepsilon^{a_1\cdots a_{N_c}}\, \varepsilon^{b_1\cdots b_{N_c}}\, \varepsilon^{c_1\cdots c_{N_c}}&=&\sum_{i,j,k=1}^{N_c}
\varepsilon^{a_1\cdots a_{i-1}b_j a_{i+1}\cdots a_{N_c}} \nn \\
&\times& \varepsilon^{b_1\cdots b_{j-1}c_k b_{j+1}\cdots b_{N_c}}\,  \varepsilon^{c_1\cdots c_{k-1}a_i c_{k+1}\cdots c_{N_c}}, 
\eeqa
from which we have
\beqa
3 [BBB]^{F_1,F_2,F_3}_{\Gamma_1,\Gamma_2,\Gamma_3} -
\sum_{i,j,k=1}^3[ BBB]^{(F_1,F_2,F_3)[i,j,k]}_{(\Gamma_1,\Gamma_2,\Gamma_3)[i,j,k]} &=&0
\label{gaugeid3}
\eeqa
where the $i$-th index of $ABC$, the $j$-th index of $DEF$ and the $k$-th index of $GHI$ are cyclically interchanged in $(ABC,DEF,GHI)[i,j,k]$. For example, 
\beqa
(\Gamma_1,\Gamma_2,\Gamma_3)[1,1,2] &=&
(\beta_3\beta_1\gamma_1,\alpha_1\beta_2\gamma_2,\alpha_3\alpha_2\gamma_3), \nonumber \\
(\Gamma_1,\Gamma_2,\Gamma_3)[1,2,3] &=&
(\gamma_3\beta_1\gamma_1,\alpha_2\alpha_1\gamma_2,\alpha_3\beta_3\beta_2). \nonumber
\eeqa
Note that the cyclic interchange of indices occurs  simultaneously for both $\Gamma_i$ and $F_i$ in the above formula. Both 2N and 3N identities are incorporated in our calculation.

\subsection{Results of anomalous dimensions for 9 quark operators at 1-loop}
In tables \ref{5f4gB}--\ref{5f4gH}, we give all eigenvalues of the matrix 
$\gamma_{AB}^{(1)}$ 
in units of $2d$, which were calculated and checked independently
by Mathematica and Maple programs, 
for $F_1\cup F_2\cup F_3 = (fffffgggg)$ with $f\not=g$ and all independent 
combinations of 
$\Gamma_1\cup \Gamma_2\cup \Gamma_3$. 
The five digits 
$n_1,n_3,n_5,n_7,n_9$  in the isospin column give the number of representations
with isospins $1/2,3/2,5/2,7/2,9/2$, respectively. For example, $32100$ means
3 operators with $I=1/2$, 2 with $I=3/2$, 1 with $I=5/2$ and 0 
with $I=7/2, 9/2$. 

The results in the tables show some notable patterns.
Firstly the eigenvalues $\gamma_j/2d$ are all even integers;
this is non-trivial since there appears to be considerable 
mixing in our initial operator bases. 
Secondly there is a tendency for the operators with
larger isospin to have smaller (more negative) eigenvalues.
Thirdly there are relations between the entries in the tables 
for different Dirac indices; e.g. the isospin degeneracies 
for the indices 111113344 and 111122223 in table 1 are identical
and furthermore all the corresponding eigenvalues related 
by a common shift of $-32$. 
These observations suggest that there is an underlying 
algebraic structure that we have unfortunately not
yet been able to reveal.
\newpage

\noindent
Note that a combination obtained from an other one  by  

i) the interchange $(1,2)\leftrightarrow (3,4)$ or 

ii) the simultaneous interchange of $ 1\leftrightarrow2$ and
$3\leftrightarrow4$ or 

iii) the interchange $1\leftrightarrow2$

\noindent
has obviously the same spectrum of anomalous dimensions and for this
reason not listed separately. 

The star symbol ${}^*$ next to the eigenvalues means that there is a 
corresponding operator which overlaps with the tree operator in 
eq. (\ref{eq:local3N}). Since the tree operator has $I=1/2$, this can only 
happen if the corresponding $n_1$ is different from zero. The tree operator
is invariant under the symmetry i) above whereas its two spin components are
exchanged under the symmetry ii). This means that whenever the tree operator
overlaps with a particular operator, it also overlaps with these transformed 
ones. The meaning of the symbol ${}^\natural$ is an overlap between the tree
operator and the transform of the operator under symmetry iii).
It is intriguing that the tree operator generally overlaps
with the largest eigenvalue for given Dirac indices;
again a fact for which we do not yet have a simple explanation.

As a simple example, let us consider the Dirac index distribution 111112222 
(third entry in table 1). The part of the tree operator relevant for this case
is
\begin{equation}
T_1=(B^{ffg}_{112}-B^{ffg}_{121})\,
(B^{ffg}_{212}-B^{ffg}_{221})\,
(B^{gfg}_{112}-B^{gfg}_{121})\,.
\end{equation} 
There are altogether 53 local 9-quark operators with this distribution of
Dirac indices, but the number of independent ones is reduced to 2 after imposing
all the gauge identities (\ref{gaugeid2}) and (\ref{gaugeid3}). A possible 
choice is
\begin{equation}
O_1=B^{fff}_{111}\,B^{ffg}_{221}\,B^{ggg}_{122},\qquad\quad
O_2=B^{fff}_{112}\,B^{ffg}_{221}\,B^{ggg}_{112}.\qquad\quad
\end{equation}
Using the gauge identities we have
\begin{equation}
T_1=\frac{5}{6}O_1-5O_2,
\end{equation}
which is proportional to the operator (of $I=1/2$) corresponding to 
anomalous dimension -60 in our units. The other combination
\begin{equation}
O_1+3O_2
\end{equation}
has anomalous dimension -84 and $I=3/2$ and has no overlap with $T_1$.
As explained above, among the operators with Dirac index distribution
333334444 there is one with anomalous dimension -60 which also overlaps with
the spin=1 component of the tree operator and among the ones corresponding
to 222221111 one which overlaps with its spin=2 component. The space of
operators corresponding to other Dirac index distributions are considerably 
larger than in this example. For example, the case 111223344 (last entry in
table 4) has 1369 operators before, and 117 operators after imposing the 
gauge identities.

As can be seen from  tables \ref{5f4gB}--\ref{5f4gH}, the largest eigenvalue of 
$\gamma_{AB}^{(1)}$ is $16d$ (occurring already at the tree level). Therefore, the 
largest eigenvalue 
of $\tilde\gamma_{AB}^{(1)}$ becomes $2d\times (8-36) $, which is negative, so 
that $\beta_A^{\rm tree} = -14/(33-2N_f)$.
Therefore, in conclusion,  the operators at the tree level in OPE dominate 
at short distance in the 3N NBS wave function.

\section{Short distance repulsion of 3NF }
\label{sec:discussion}
As discussed before, the 3N potential at short distance is given by
\beqa
V_{2N+3NF}(\vec r,\vec\rho) \simeq \frac{ 1}{m_N} 
\frac{-4\beta_A^{\rm tree}}{s^2(-\log(\Lambda s))} ,
\eeqa
where $\beta_A^{\rm tree}$ is given in eq.~(\ref{eq:betaA}).
Since this result dominates over the one appearing in the 2N potential,
which is of the form (\ref{nu2}), 
the above behavior of 
$V_{2N+3NF}(\vec r,\vec\rho) $ at short distance must come solely from 
$V_{3NF}(\vec r,\vec\rho)$.
Unlike for the 2NF no additional nonperturbative considerations are required
in this case and
 therefore we can conclude that the 3NF is always repulsive at short distance, 
which is universal in the sense that it does not depend on the details of the 
3N state, used to define the NBS  wave function, such as its energy $E$.
Note however that this conclusion is valid for the 3NF in our definition 
({\it i.e.} defined from the NBS wave function) since potentials are not 
observable in general and therefore scheme-dependent.
Unless one fixes the scheme for the definition of the potential , it is 
meaningless to ask whether the 3NF has a repulsive core or not.  By using the 
potential scheme considered in this paper, we can show that the 3NF universally
have repulsive cores.  

Finally we would like to note that as listed in Tables 1-4,
the total number of 9-quark local operators is several hundred and the
spectrum
of anomalous dimensions is rather dense. We singled out the ones with largest
anomalous dimensions which dominate at short distance but it is not evident
at what distance scale this leading behavior sets in. Even if it turns out
that
these individual operators really dominate at extremely short distances only,
our main conclusion may remain valid due to the fact that all the eigenvalues
of the effective gamma matrix are negative (corresponding to short distance
repulsion). We think a simple explanation of this fact should exist
(maybe related to the Pauli principle).

An interesting and important extension of the present analysis is to 
investigate the short distance behavior of the three baryon force (3BF) by the 
same method. Its results can tell us whether there is a universal short 
distance repulsion also in the 3BF, which has been suggested to explain the 
observed maximum mass of neutron stars \cite{Nishizaki:2002ih}.
    

\section*{Acknowledgments}
S.~A would like to thank Dr. T.~Doi, Prof. T.~Hatsuda, Dr. N.~Ishii for useful 
discussions.
S.~A. is supported in part by  Grant-in-Aid for Scientific Research on 
Innovative Areas (No. 2004: 20105001,20105003) and by SPIRE(Strategic Program for Innovative REsearch).
This investigation was also supported in part by the Hungarian National
Science Fund OTKA (under K83267).
S.~A. and J.~B. would like to thank the Max-Planck-Institut f\"ur Physik 
for its kind hospitality during their stay for this research project.


\clearpage
\begin{table}[h] 
\centering 
\begin{tabular}[t]{c|c|c} 
\hline 
Dirac indices&$\gamma_j/(2d)$& isospins\\[1.0ex] 
\hline  
\hline  
$ 111111222 $ & $-84$ & $01000$\\[1.0ex] 
\hline  
$ 111111333 $ & $-36$ & $01000$\\[1.0ex] 
\hline  
$ 111112222 $ & $-84$ & $01000$\\[1.0ex] 
$$ & $-60^*$ & $10000$\\[1.0ex] 
\hline  
$ 111113333 $ & $-28$ & $11000$\\[1.0ex] 
\hline  
$ 111111223 $ & $-60$ & $11000$\\[1.0ex] 
\hline  
$ 111111334 $ & $-36$ & $01000$\\[1.0ex] 
$$ & $-12$ & $10000$\\[1.0ex] 
\hline  
$ 111111332 $ & $-44$ & $11000$\\[1.0ex] 
\hline  
$ 111112223 $ & $-76$ & $01100$\\[1.0ex] 
$$ & $-60$ & $11000$\\[1.0ex] 
$$ & $-52$ & $11000$\\[1.0ex] 
$$ & $-40^\natural$ & $10000$\\[1.0ex] 
\hline  
$ 111113334 $ & $-44$ & $01100$\\[1.0ex] 
$$ & $-28$ & $11000$\\[1.0ex] 
$$ & $-20$ & $11000$\\[1.0ex] 
$$ & $-8$ & $10000$\\[1.0ex] 
\hline  
$ 111113332 $ & $-36$ & $23100$\\[1.0ex] 
$$ & $-24$ & $10000$\\[1.0ex] 
\hline  
$ 111112233 $ & $-52$ & $12100$\\[1.0ex] 
$$ & $-44$ & $11000$\\[1.0ex] 
$$ & $-34$ & $21000$\\[1.0ex] 
\hline  
$ 111113344 $ & $-44$ & $01100$\\[1.0ex] 
$$ & $-28$ & $11000$\\[1.0ex] 
$$ & $-20$ & $11000$\\[1.0ex] 
$$ & $-16$ & $11000$\\[1.0ex] 
$$ & $-8$ & $10000$\\[1.0ex] 
\hline  
$ 111133332 $ & $-36$ & $12100$\\[1.0ex] 
$$ & $-28$ & $11000$\\[1.0ex] 
$$ & $-18$ & $21000$\\[1.0ex] 
\hline  
\end{tabular} 
$\phantom{\pi\pi\pi\pi\pi\pi}$
\begin{tabular}[t]{c|c|c} 
\hline 
Dirac indices&$\gamma_j/(2d)$& isospins\\[1.0ex] 
\hline 
\hline  
$ 111122223 $ & $-76$ & $01100$\\[1.0ex] 
$$ & $-60$ & $11000$\\[1.0ex] 
$$ & $-52$ & $11000$\\[1.0ex] 
$$ & $-48^*$ & $11000$\\[1.0ex] 
$$ & $-40^*$ & $10000$\\[1.0ex] 
\hline  
$ 111122233 $ & $-76$ & $01110$\\[1.0ex] 
$$ & $-52$ & $12100$\\[1.0ex] 
$$ & $-46$ & $12100$\\[1.0ex] 
$$ & $-44$ & $11000$\\[1.0ex] 
$$ & $-34$ & $21000$\\[1.0ex] 
$$ & $-28^\natural$ & $21000$\\[1.0ex] 
\hline  
$ 111133344 $ & $-60$ & $01110$\\[1.0ex] 
$$ & $-36$ & $12100$\\[1.0ex] 
$$ & $-30$ & $12100$\\[1.0ex] 
$$ & $-28$ & $11000$\\[1.0ex] 
$$ & $-18$ & $21000$\\[1.0ex] 
$$ & $-12$ & $21000$\\[1.0ex] 
\hline  
$ 111133322 $ & $-52$ & $12210$\\[1.0ex] 
$$ & $-36$ & $23100$\\[1.0ex] 
$$ & $-28$ & $22000$\\[1.0ex] 
$$ & $-24$ & $10000$\\[1.0ex] 
$$ & $-16$ & $11000$\\[1.0ex] 
\hline  
$ 111222333 $ & $-84$ & $01111$\\[1.0ex] 
$$ & $-52$ & $12210$\\[1.0ex] 
$$ & $-48$ & $01100$\\[1.0ex] 
$$ & $-36$ & $23100$\\[1.0ex] 
$$ & $-28$ & $22000$\\[1.0ex] 
$$ & $-24$ & $32100$\\[1.0ex] 
$$ & $-16$ & $11000$\\[1.0ex] 
\hline  
$ 111111234 $ & $-44$ & $11000$\\[1.0ex] 
$$ & $-28$ & $10000$\\[1.0ex] 
\hline  
\end{tabular} 
\caption{\footnotesize Eigenvalues $\gamma_j$ of the anomalous
dimension matrix $\gamma$ and isospins of the corresponding eigenvectors 
for the case 5f4g. The five digits $n1,n3,n5,n7,n9$ in the isospins column
give the number of representations with isospins $1/2,3/2,5/2,7/2,9/2$,
respectively.}  
\label{5f4gB} 
\end{table}

\clearpage



\begin{table}[h] 
\centering 
\begin{tabular}[t]{c|c|c} 
\hline 
Dirac indices&$\gamma_j/(2d)$& isospins\\[1.0ex] 
\hline 
\hline  
$ 111112234 $ & $-60$ & $11100$\\[1.0ex] 
$$ & $-52$ & $12100$\\[1.0ex] 
$$ & $-44$ & $11000$\\[1.0ex] 
$$ & $-42$ & $11000$\\[1.0ex] 
$$ & $-36$ & $01000$\\[1.0ex] 
$$ & $-34$ & $21000$\\[1.0ex] 
$$ & $-28$ & $10000$\\[1.0ex] 
$$ & $-18$ & $10000$\\[1.0ex] 
\hline  
$ 111113324 $ & $-48$ & $11100$\\[1.0ex] 
$$ & $-36$ & $24100$\\[1.0ex] 
$$ & $-30$ & $11000$\\[1.0ex] 
$$ & $-24$ & $10000$\\[1.0ex] 
$$ & $-18$ & $10000$\\[1.0ex] 
$$ & $-12$ & $21000$\\[1.0ex] 
\hline  
$ 111122234 $ & $-76$ & $01110$\\[1.0ex] 
$$ & $-60$ & $11200$\\[1.0ex] 
$$ & $-54$ & $11100$\\[1.0ex] 
$$ & $-52$ & $12100$\\[1.0ex] 
$$ & $-46$ & $12100$\\[1.0ex] 
$$ & $-44$ & $11000$\\[1.0ex] 
$$ & $-42$ & $11000$\\[1.0ex] 
$$ & $-36^*$ & $12000$\\[1.0ex] 
$$ & $-34$ & $21000$\\[1.0ex] 
$$ & $-30$ & $01000$\\[1.0ex] 
$$ & $-28^*$ & $31000$\\[1.0ex] 
$$ & $-18$ & $10000$\\[1.0ex] 
$$ & $-12^*$ & $10000$\\[1.0ex] 
\hline  
\end{tabular} 
$\phantom{\pi\pi\pi\pi\pi\pi}$
\begin{tabular}[t]{c|c|c} 
\hline 
Dirac indices&$\gamma_j/(2d)$& isospins\\[1.0ex] 
\hline 
\hline  
$ 111133324 $ & $-52$ & $11110$\\[1.0ex] 
$$ & $-44$ & $01100$\\[1.0ex] 
$$ & $-40$ & $11100$\\[1.0ex] 
$$ & $-36$ & $12100$\\[1.0ex] 
$$ & $-34$ & $12100$\\[1.0ex] 
$$ & $-28$ & $22100$\\[1.0ex] 
$$ & $-22$ & $21000$\\[1.0ex] 
$$ & $-20$ & $11000$\\[1.0ex] 
$$ & $-18$ & $21000$\\[1.0ex] 
$$ & $-16$ & $01000$\\[1.0ex] 
$$ & $-10$ & $11000$\\[1.0ex] 
$$ & $-8$ & $10000$\\[1.0ex] 
$$ & $2$ & $10000$\\[1.0ex] 
\hline  
$ 111122334 $ & $-64$ & $11110$\\[1.0ex] 
$$ & $-52$ & $12210$\\[1.0ex] 
$$ & $-48$ & $11100$\\[1.0ex] 
$$ & $-46$ & $01100$\\[1.0ex] 
$$ & $-40$ & $33200$\\[1.0ex] 
$$ & $-36$ & $24100$\\[1.0ex] 
$$ & $-30$ & $11000$\\[1.0ex] 
$$ & $-28$ & $24100$\\[1.0ex] 
$$ & $-24$ & $10000$\\[1.0ex] 
$$ & $-22$ & $22000$\\[1.0ex] 
$$ & $-18$ & $10000$\\[1.0ex] 
$$ & $-16$ & $11000$\\[1.0ex] 
$$ & $-12$ & $21000$\\[1.0ex] 
$$ & $-10^\natural$ & $10000$\\[1.0ex] 
$$ & $8^\natural$ & $10000$\\[1.0ex] 
\hline

\end{tabular} 
\caption{\footnotesize As in Table~1 (continued).}  
\label{5f4gD} 
\end{table}

\clearpage


\begin{table}[h] 
\centering 
\begin{tabular}[t]{c|c|c} 
\hline 
Dirac indices&$\gamma_j/(2d)$& isospins\\[1.0ex] 
\hline 
\hline  
$ 111133442 $ & $-60$ & $11110$\\[1.0ex] 
$$ & $-52$ & $11110$\\[1.0ex] 
$$ & $-44$ & $01100$\\[1.0ex] 
$$ & $-42$ & $01100$\\[1.0ex] 
$$ & $-40$ & $11100$\\[1.0ex] 
$$ & $-36$ & $12100$\\[1.0ex] 
$$ & $-34$ & $12100$\\[1.0ex] 
$$ & $-30$ & $11000$\\[1.0ex] 
$$ & $-28$ & $22100$\\[1.0ex] 
$$ & $-24$ & $12100$\\[1.0ex] 
$$ & $-22$ & $21000$\\[1.0ex] 
$$ & $-20$ & $11000$\\[1.0ex] 
$$ & $-18$ & $21000$\\[1.0ex] 
$$ & $-16$ & $12000$\\[1.0ex] 
$$ & $-10$ & $11000$\\[1.0ex] 
$$ & $-8$ & $10000$\\[1.0ex] 
$$ & $-6$ & $21000$\\[1.0ex] 
$$ & $2$ & $10000$\\[1.0ex] 
\hline  
\end{tabular} 
$\phantom{\pi\pi\pi\pi\pi\pi}$
\begin{tabular}[t]{c|c|c} 
\hline 
Dirac indices&$\gamma_j/(2d)$& isospins\\[1.0ex] 
\hline 
\hline  
$ 111222334 $ & $-84$ & $01111$\\[1.0ex] 
$$ & $-64$ & $11110$\\[1.0ex] 
$$ & $-60$ & $11220$\\[1.0ex] 
$$ & $-52$ & $12210$\\[1.0ex] 
$$ & $-48$ & $12200$\\[1.0ex] 
$$ & $-46$ & $01100$\\[1.0ex] 
$$ & $-42$ & $01100$\\[1.0ex] 
$$ & $-40$ & $33200$\\[1.0ex] 
$$ & $-36$ & $35200$\\[1.0ex] 
$$ & $-30$ & $11000$\\[1.0ex] 
$$ & $-28$ & $24100$\\[1.0ex] 
$$ & $-24^*$ & $42100$\\[1.0ex] 
$$ & $-22$ & $22000$\\[1.0ex] 
$$ & $-18^*$ & $21000$\\[1.0ex] 
$$ & $-16$ & $11000$\\[1.0ex] 
$$ & $-12$ & $21000$\\[1.0ex] 
$$ & $-10^*$ & $10000$\\[1.0ex] 
$$ & $0^*$ & $11000$\\[1.0ex] 
$$ & $8^*$ & $10000$\\[1.0ex] 
\hline  
\end{tabular} 
\caption{\footnotesize As in Table~1 (continued).}  
\label{5f4gF} 
\end{table}

\clearpage

\begin{table}[h] 
\centering 
\begin{tabular}[t]{c|c|c} 
\hline 
Dirac indices&$\gamma_j/(2d)$& isospins\\[1.0ex] 
\hline 
\hline  
$ 111333224 $ & $-76$ & $11111$\\[1.0ex] 
$$ & $-60$ & $01110$\\[1.0ex] 
$$ & $-52$ & $12220$\\[1.0ex] 
$$ & $-46$ & $22210$\\[1.0ex] 
$$ & $-44$ & $01100$\\[1.0ex] 
$$ & $-40$ & $11200$\\[1.0ex] 
$$ & $-36$ & $12100$\\[1.0ex] 
$$ & $-34$ & $24200$\\[1.0ex] 
$$ & $-30$ & $12100$\\[1.0ex] 
$$ & $-28$ & $34200$\\[1.0ex] 
$$ & $-22$ & $32100$\\[1.0ex] 
$$ & $-20$ & $11000$\\[1.0ex] 
$$ & $-18$ & $21000$\\[1.0ex] 
$$ & $-16^\natural$ & $22000$\\[1.0ex] 
$$ & $-12$ & $21000$\\[1.0ex] 
$$ & $-10$ & $12000$\\[1.0ex] 
$$ & $-8$ & $10000$\\[1.0ex] 
$$ & $-4$ & $11000$\\[1.0ex] 
$$ & $2$ & $10000$\\[1.0ex] 
$$ & $8^\natural$ & $10000$\\[1.0ex] 
\hline  
\end{tabular} 
$\phantom{\pi\pi\pi\pi\pi\pi}$
\begin{tabular}[t]{c|c|c} 
\hline 
Dirac indices&$\gamma_j/(2d)$& isospins\\[1.0ex] 
\hline 
\hline  
$ 111223344 $ & $-76$ & $11111$\\[1.0ex] 
$$ & $-60$ & $12220$\\[1.0ex] 
$$ & $-54$ & $11110$\\[1.0ex] 
$$ & $-52$ & $12220$\\[1.0ex] 
$$ & $-48$ & $01110$\\[1.0ex] 
$$ & $-46$ & $22210$\\[1.0ex] 
$$ & $-44$ & $01100$\\[1.0ex] 
$$ & $-42$ & $12200$\\[1.0ex] 
$$ & $-40$ & $11200$\\[1.0ex] 
$$ & $-36$ & $13200$\\[1.0ex] 
$$ & $-34$ & $24200$\\[1.0ex] 
$$ & $-30$ & $23100$\\[1.0ex] 
$$ & $-28$ & $34200$\\[1.0ex] 
$$ & $-24^*$ & $23100$\\[1.0ex] 
$$ & $-22$ & $32100$\\[1.0ex] 
$$ & $-20$ & $11000$\\[1.0ex] 
$$ & $-18$ & $33100$\\[1.0ex] 
$$ & $-16^*$ & $33000$\\[1.0ex] 
$$ & $-12$ & $21000$\\[1.0ex] 
$$ & $-10$ & $12000$\\[1.0ex] 
$$ & $-8$ & $10000$\\[1.0ex] 
$$ & $-6$ & $21000$\\[1.0ex] 
$$ & $-4$ & $11000$\\[1.0ex] 
$$ & $0^*$ & $21000$\\[1.0ex] 
$$ & $2$ & $10000$\\[1.0ex] 
$$ & $8^*$ & $10000$\\[1.0ex] 
\hline  
\end{tabular} 
\caption{\footnotesize As in Table~1 (continued).}  
\label{5f4gH} 
\end{table}

\end{document}